\documentclass[aps,amsmath,amssymb,10pt]{revtex4}
\usepackage{graphicx}
\usepackage[T1]{fontenc}
\usepackage{parskip}

\newcommand{\dd}{\mathrm{d}}
\newcommand{\ee}{\mathrm{e}}

\begin{document}

\title{On the relevance of avoided crossings away from quantum
  critical point to the complexity of quantum adiabatic algorithm.}
\affiliation{Applied Physics Center at NASA Ames Research Center,  Moffett Field, CA 94035-1000.}
\author{Sergey Knysh}
\email{Sergey.I.Knysh@nasa.gov}
\affiliation{Applied Physics Center at NASA Ames Research Center,  Moffett Field, CA 94035-1000.}
\author{Vadim Smelyanskiy}
\email{Vadim.N.Smelyanskiy@nasa.gov}
\affiliation{Applied Physics Center at NASA Ames Research Center,  Moffett Field, CA 94035-1000.}
\date{\today}

\begin{abstract}
  Two recent preprints [B.~Altshuler, H.~Krovi, and J.~Roland, ``Quantum adiabatic optimization fails for 
  random instances of NP-complete problems'', arXiv:0908.2782 and ``Anderson localization casts clouds 
  over adiabatic quantum optimization'', arXiv:0912.0746] argue that random 4th order perturbative corrections
  to the energies of local minima of random instances of NP-complete problem lead to avoided crossings that
  cause the failure of quantum adiabatic algorithm (due to exponentially small gap) close to the end, for very
  small transverse field that scales as an inverse power of instance size $N$. The theoretical portion of this work 
  does not to take into account the exponential degeneracy of the ground and excited states at zero field. 
  A corrected analysis shows that unlike those in the middle of the spectrum, avoided crossings at the \emph{edge}
  would require high [$O(1)$] transverse fields, at which point the perturbation theory may become divergent 
  due to quantum phase transition. This effect manifests itself only in large instances [$\exp(0.02 N) \gg 1$],
  which might be the reason it had not been observed in the authors' numerical work. While we dispute the 
  proposed mechanism of failure of quantum adiabatic algorithm, we cannot draw any conclusions on its
  ultimate complexity.
\end{abstract}

\maketitle

\textbf{Quantum adiabatic algorithm.} The quantum adiabatic algorithm is a
generic algorithm proposed for the solution of a variety of combinatorial
optimization and decision problems involving binary variables from the
NP-complete family. One considers a Hamiltonian involving $N$ qubits $x_i$ (or
$N$ spins) dependent on parameter $\lambda$:
\begin{equation}
  \hat{H} (\lambda) = \sum_{\boldsymbol{x} \in \{0, 1\}^N} E (\boldsymbol{x})
  |\boldsymbol{x} \rangle \langle \boldsymbol{x}| - \lambda \sum_k
  \sum_{\boldsymbol{x} \in \{0, 1\}^N} |x_1 \ldots x_k \ldots x_N \rangle
  \langle x_1 \ldots \bar{x}_k \ldots x_N |, \label{Hdef}
\end{equation}
where $\bar{x}_k = 1 - x_k$ represents qubit flip. Here $E (\boldsymbol{x})$ is
the cost function corresponding to bit assignment $\boldsymbol{x}=(x_1, \ldots,x_N)$.
A solution is verifiable in polynomial time (but finding one may take
exponential time) hence the Hamiltonian is implementable using only polynomially
large number of gadgets. The system is initially prepared in a state that is
the symmetric superposition of all $2^N$ possible bit assignments ---
an exact ground state of the second (``driver'') term in (\ref{Hdef}),
which corresponds to the uniform magnetic field $\lambda$ in the direction
orthogonal to  the quantization axis of computational basis $|\boldsymbol{x} \rangle$. 
The parameter $\lambda$ is changed in time from $\lambda (0) \gg 1$
initially to $\lambda (T) = 0$ at the end of the algorithm. By adiabatic
theorem, the system will remain in its ground state with high probability
provided that $\dd \lambda / \dd t \ll \Delta^2(\lambda)$, where $\Delta
(\lambda) = E_1 (\lambda) - E_0 (\lambda)$ is the energy gap between the
ground state of $\hat{H} (\lambda)$ and its first excited state. At time $t =
T$ the system will be in a superposition state of configurations with the
optimal cost and one of the optimal solutions may be obtained by preforming a
final measurement on the qubits. The running time of the algorithm (its
complexity) is given by $T \sim 1 / \Delta_{\min}^2$, where $\Delta_{\min}$ is
the minimum value of the gap as a function of $\lambda$. The value
$\lambda=\lambda_{\ast}$ for which the gap is minimal will be referred to as the
bottleneck of the algorithm.

It is known that the minimum gap can be exponentially small in $N$ in the
worst case. A really interesting question is how adiabatic algorithm performs
on random instances of combinatorial optimization problems --- the
typical-case complexity. Historically, the benchmark problem for the quantum
adiabatic algorithm has been the exact cover problem. An instance of
random exact cover problem is a set of $N$ bits and $M$ clauses, each clause
$C$ containing three bits $(x_{i_C}, x_{j_C}, x_{k_C})$ chosen uniformly at
random. One seeks an assignment such that bits in each clause add up to 1:
$x_{i_C} + x_{j_C} + x_{k_C} = 1$. A cost $(x_{i_C} + x_{j_C} + x_{k_C} - 1)^2
\geqslant 0$ is assigned to each clause so that the total cost $E
(\boldsymbol{x})$ (given by the sum over individual clauses) is zero for
\emph{satisfying assignments}. In terms of Pauli operators
[where $\hat{\sigma}_i^z |x_i \rangle = (- 1)^{x_i} |x_i \rangle$, $\hat{\sigma}_i^x
|x_i \rangle = | \bar{x}_i \rangle$], the quantum Hamiltonian is written as
(cf. Eq. (2) of Ref.~\cite{Alt09}):
\begin{equation}
  \hat{H} (\lambda) = M - \frac{1}{2} \sum_i B_i  \hat{\sigma}_i^z +
  \frac{1}{2} \sum_{\langle i j k \rangle} ( \hat{\sigma}_i^z 
  \hat{\sigma}_j^z + \hat{\sigma}_i^z  \hat{\sigma}_k^z + \hat{\sigma}_j^z 
  \hat{\sigma}_k^z) - \lambda \sum_i \hat{\sigma}_i^x, \label{Hl}
\end{equation}
where $B_i$ is the number of clauses in which bit $i$ appears and the sum in
the third term is over all clauses $\langle i j k \rangle$. The first three
terms describe the problem Hamiltonian that is diagonal in
$\hat{\sigma}^z$-representation, while the fourth term describes a
magnetic field in the transverse direction.

It has been observed that the bottleneck of simulated annealing (which can be
thought of as a classical counterpart of quantum adiabatic algorithm) is the
vicinity of temperature-driven phase transition. One might conjecture that the
bottleneck of quantum adiabatic algorithm is the vicinity of
transverse-field--driven quantum phase transition at finite $\lambda =
\lambda_c > 0$. This indeed had been confirmed in a few random NP-complete
problems \cite{QPT}. However, there is no reason to expect that this scenario is
universal; even the existence of the phase transition cannot be guaranteed in
some models \cite{noQPT}. Ref.~\cite{Alt09} asserts that the bottleneck of the quantum
adiabatic algorithm for random exact cover is unconnected to the quantum phase
transition but is due to ``accidental'' avoided crossings of energy levels
corresponding to localized states for infinitesimal transverse fields
($\lambda \rightarrow 0$ as $N \rightarrow \infty$). The associated gap is
related to the overlap between localized states and is expected to be
exponentially small. It is claimed that this mechanism is not peculiar to the
random exact cover problem, but applies to a wide range of NP-complete
problems defined on random hypergraphs. The possibility of avoided crossings
for $\lambda < \lambda_c$ had been raised before, for a model with a
quasi-continuous (level spacings $\ll 1$) spectrum \cite{Santoro02}, but was thought 
not to occur for models with a discrete spectrum, such as exact cover, K-SAT, etc. 
Ref.~\cite{Alt09} predicts avoided crossings close to the end of the algorithm whereas 
recent quantum Monte Carlo (QMC) simulations show the bottleneck in the middle of the 
algorithm \cite{QMC}. However, QMC studies consider a different ensemble (extremely rare
instances with a unique satisfying assignment are chosen) and the problem sizes considered
($N = 256$) may be too small (Ref.~\cite{Alt09} estimates that the described mechanism may
not kick in until $N \sim 10^5$). We will demonstrate that the exponential
degeneracy of the ground state, which is a distinguishing feature of random
NP-complete problems with discrete spectrum addressed in \cite{Alt09}, dooms the
proposed mechanism. Note that when the instance is not drawn from a uniformly
random ensemble but is instead crafted to contain exactly one global and one
local minimum separated by $N$ bit flips, the avoided crossing does take place
for $\lambda \rightarrow 0$ \cite{Farhi09}.

\textbf{Overview of perturbation theory analysis.} Ref.~\cite{Alt09} starts 
with the classical Hamiltonian for an instance with $N$ bits and $M$ clauses and
develops a perturbation theory in a small parameter $\lambda \ll 1$. In this
limit the perturbation theory is expected to be \emph{locally convergent}.
One may consider a global minimum $E (\boldsymbol{x}_0) = 0$ and a local minimum
at $E (\boldsymbol{x}_1) = 1$. For small $\lambda > 0$, these energy levels
acquire perturbative corrections
\begin{equation}
  E_{\boldsymbol{x}_0}^{(M)} (\lambda) \approx \delta E_{\boldsymbol{x}_0}^{(M)}
  (\lambda), \hspace{2em} E_{\boldsymbol{x}_1}^{(M)} (\lambda) \approx 1 + \delta
  E_{\boldsymbol{x}_1}^{(M)} (\lambda) .
\end{equation}
For some value $\lambda$ such that $\delta E_{\boldsymbol{x}_0}^{(M)}(\lambda) - 
\delta E_{\boldsymbol{x}_1}^{(M)} (\lambda) = 1$, the levels
corresponding to states localized near $\boldsymbol{x} = \boldsymbol{x}_0$
and $\boldsymbol{x} = \boldsymbol{x}_1$ will be equal in energy. The minimum gap
will be non-zero, but exponentially small, provided that $\boldsymbol{x}_0$ and
$\boldsymbol{x}_1$ differ by $O(N)$ bit flips.

\begin{figure}[!ht]
\includegraphics{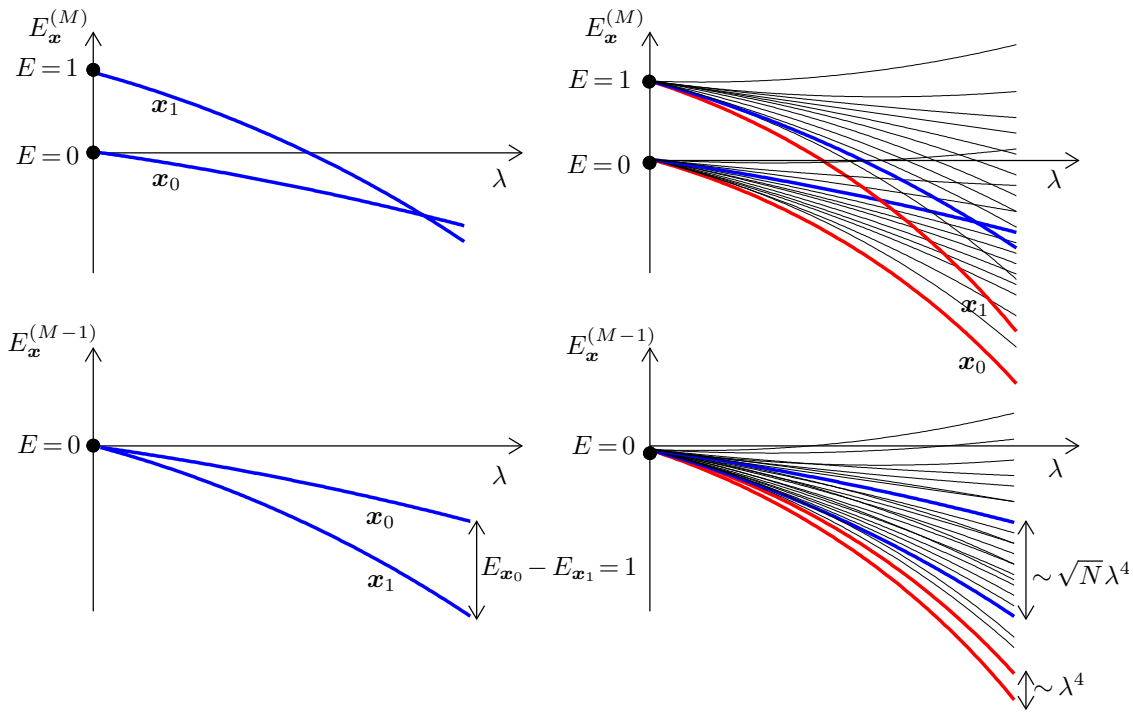}
\caption{\emph{Left:} Top figure shows the intersection of energy levels corresponding
to a global [$E(\boldsymbol{x}_0)=0$] and a local [$E(\boldsymbol{x}_1)=1$] minimum. 
Bottom figure shows two levels corresponding to global minima for an instance with one 
less clause. When the level splitting equals one at $\lambda=\lambda_\ast$, the levels 
of the original instance will cross for somewhat smaller $\lambda$.
\emph{Right:} The effect of exponential degeneracy is depicted here. Level splitting
(bottom figure) is large only for randomly chosen levels (blue lines) so they will
intersect (top figure) with high probability. Level with the smallest perturbation
correction ($\boldsymbol{x}_0$ and $\boldsymbol{x}_1$) may not intersect until $\lambda$
is large as the level splitting does not scale with $N$.
\label{fig:crossing}}
\end{figure}

In a drastic simplification, Ref.~\cite{Alt09} uses a clever trick to show that It suffices
to examine only the global minima.
Once a clause contradicting $\boldsymbol{x}_1$ is removed, both $\boldsymbol{x}_0$
and $\boldsymbol{x}_1$ will have zero cost. Writing $E^{(M - 1)}_{\boldsymbol{x}}
(\lambda)$ for the energy of eigenstate localized near $\boldsymbol{x}$ for the
instance with $M - 1$ clauses, and $\lambda_{\ast}$ denoting the solution to
\begin{equation}
  E_{\boldsymbol{x}_0}^{(M - 1)} (\lambda_{\ast}) - E_{\boldsymbol{x}_1}^{(M - 1)}
  (\lambda_{\ast}) = 1, \label{DEM1}
\end{equation}
it can be argued that an instance with $M$ clauses should have an avoided
crossing for some $\lambda < \lambda_{\ast}$. This follows from inequality
satisfied for small $\lambda$,
\begin{equation}
  0 < E_{\boldsymbol{x}}^{(M)} (\lambda) - E_{\boldsymbol{x}}^{(M-1)}(\lambda) < 1,
\label{MM1}
\end{equation}
obtained by treating the $M$-th clause as a perturbation. Eqs. (\ref{DEM1}) and
(\ref{MM1}) together imply $E_{\boldsymbol{x}_0}^{(M)} (\lambda_{\ast}) >
E_{\boldsymbol{x}_1}^{(M)} (\lambda_{\ast})$, but the opposite inequality holds
for $\lambda = 0$. Therefore, the energies, being continuous functions of
$\lambda$, must be equal for some $\lambda < \lambda_{\ast}$. This
construction is visualized in Fig. \ref{fig:crossing} (left).

In general, one considers a random instance with $M - 1$ clauses chosen
uniformly at random and some solution $\boldsymbol{x}_1$. A new uniformly random
instance with $M$ clauses is formed by adding a new random clause. With
\emph{finite probability}, the new clause is violated by $\boldsymbol{x}_1$ so
that levels corresponding to $\boldsymbol{x}_1$ and some other solution
$\boldsymbol{x}_0$ satisfied by the new clause cross for $\lambda$ such that
$\Delta E_{10} = E_{\boldsymbol{x}_0}^{(M-1)} (\lambda) - E_{\boldsymbol{x}_1}^{(M-1)} (\lambda) \sim 1$.
Failing that, random instances with $M + 1$, $M + 2$, etc. clauses may be generated by
adding more random clauses, which ensures that avoided crossing takes place
with probability tending to one.

The necessary condition for this mechanism is the convergence of the perturbation
theory. Ref.~\cite{Alt09} justifies its use by showing that $\lambda_{\ast}
\rightarrow 0$ as $N \rightarrow \infty$. Within ordinary (non-degenerate) perturbation theory 
up to the 4th order, the energy of the state corresponding to a
solution with zero cost is (cf. Eq.~(33) in Ref.~\cite{Alt09})
\begin{equation}
  E_{\boldsymbol{x}} (\lambda) \approx \text{\emph{common term}} + \lambda^4
  \sum_{\langle i j k \rangle} \left( \frac{4 / (B_j B_k)^2}{1 - 4 / (B_j +
  B_k)^2} x_i + \frac{4 / (B_i B_k)^2}{1 - 4 / (B_i + B_k)^2} x_j + \frac{4 /
  (B_i B_j)^2}{1 - 4 / (B_i + B_j)^2} x_k \right) . \label{de4}
\end{equation}
The common term is the same for all configurations but the second term is
configuration-dependent. For uniformly random ensemble, $B_i$ [defined in the
text surrounding Eq. (\ref{Hl})] are random Poisson-distributed variables with
mean $3 M / N = O (1)$, each term in the sum over $M = O (N)$ clauses is a
random $O (1)$ variable. By central limit theorem, the sum is approximately a
Gaussian of width $O ( \sqrt{N})$ so that for two different bit configurations
$\Delta E_{10} \sim \sqrt{N} \lambda^4$. Therefore, Ref.~\cite{Alt09} claims that 
avoided crossings take place for $\lambda \sim 1 / N^{1 / 8} \ll 1$, well within the region 
of applicability of perturbation theory.

\textbf{Effects of exponential degeneracy.} The argument at the end of the 
previous section overlooks the fact that the values of $\boldsymbol{x}_0$ and 
$\boldsymbol{x}_1$ are correlated with realizations of random instances. While in many
circumstances neglecting correlations may not lead to qualitative changes, an
important factor in this case is the large number of solutions with zero cost.
Even if perturbative corrections to all solutions are assumed independent random
variables, it can only be established that \emph{randomly chosen} energy
levels corresponding to $E = 0$ and $E = 1$ may intersect for $\lambda \sim 1
/ N^{1 / 8}$, as depicted in Fig.~\ref{fig:crossing} (right). Since we are 
interested in the intersections with the \emph{ground state}, we require 
that $\boldsymbol{x}_1$ and $\boldsymbol{x}_0$ correspond to the ground states of
Hamiltonian with $M-1$ and $M$ clauses respectively, i.e. that
$E_{\boldsymbol{x}_1}^{(M - 1)}(\lambda)$ and $E_{\boldsymbol{x}_0}^{(M)} (\lambda)$ 
be smallest. But with this restriction, we will see that 
$E_{\boldsymbol{x}_0} (\lambda) - E_{\boldsymbol{x}_1} (\lambda) \sim \lambda^4$, 
so that avoided crossings are unlikely until $\lambda \sim 1$, which may be outside
the radius of convergence of perturbation theory.

\begin{figure}[!ht]
\begin{minipage}[b]{0.45\linewidth}
\includegraphics[width=\linewidth]{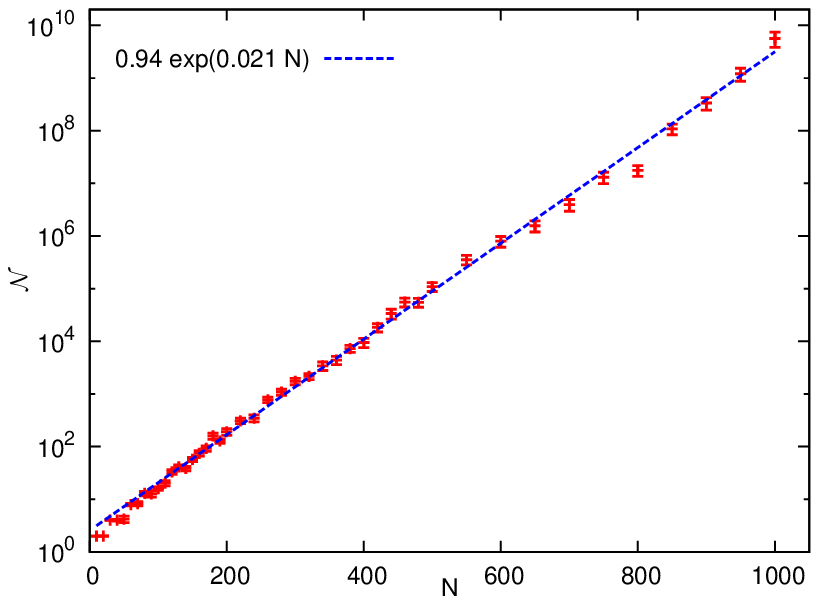}
\end{minipage}
\hspace{0.5cm}
\begin{minipage}[b]{0.45\linewidth}
\includegraphics[width=\linewidth]{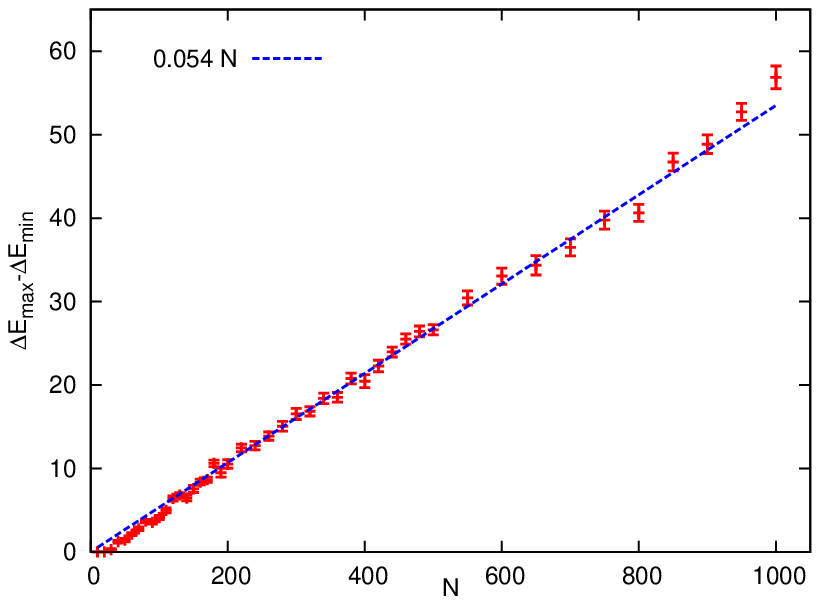}
\end{minipage}
\caption{\emph{Left:} Number of solutions (for satisfiable instances) 
as a function of $N$. The data fit exponential dependence 
$\mathcal{N} \sim \exp(cN)$ with $c \approx 0.0209 \pm 0.0002$. 
\emph{Right:} The difference between the largest and the smallest 
4th order perturbative correction to the ground state, as a function of $N$. 
A linear fit with the coefficient $0.0535 \pm 0.0004$ is obtained.
A leaf removal algorithm has been applied to randomly generated instances to
insure that no clause contains more than one bit not appearing in other clauses.
Errorbars correspond to one standard deviation (68\% confidence interval). 
Linear fits are for the interval $100 \leqslant N \leqslant 1000$.
\label{fig:degeneracy}}
\end{figure}

It is important to realize that the number of solutions of random exact cover
is exponential in $N$, even near the satisfiability threshold $\alpha_s = M /
N \approx 0.626$ where the random instance is satisfiable with probability $1/2$.
 Just prior to adding a random clause which makes an instance
unsatisfiable, some bits are frozen (have the same values in all solutions)
while others are not. The latter, \emph{``soft'' bits}, contribute to the exponential
degeneracy. In the numerical simulations of Ref.~\cite{Alt09} all bits that do
not appear in any clause as well as clauses with two or more bits that do not
belong to any other clause are removed. This ensures that flipping two bits
does not lead to another solution $E = 0$ as that would make the expression
(\ref{de4}) formally infinite, indicating that a degenerate perturbation
theory should be used instead. Such hypergraph trimming does not affect the
satisfiability of the instance, can be done in polynomial time prior to
running quantum adiabatic algorithm, and removes ``trivial degeneracies''.
However, it does not remove all degeneracies: there will remain soft bits;
moreover whether a given a given bit is soft depends on the assignment of
``hard'', or frozen bits. In Fig.~\ref{fig:degeneracy} (left) we plot the number
of solutions of trimmed hypergraph as a function of $N$ at satisfiability threshold.
It is seen that the number of solutions grows exponentially as $\mathcal{N} \approx
\exp(0.021 N)$. The smallness of exponent is the reason this effect only starts
to manifest itself for $N \gtrsim 100$.

Once the presence of exponentially many solutions [$\mathcal{N} \sim \exp (cN)$]
is taken into account, the density of states $E_{\boldsymbol{x}}(\lambda)$ for all solutions
is written as
\begin{equation}
  \rho (E) =\mathcal{N} \frac{1}{\sqrt{2 \pi N} \sigma} \exp
  \left[ - \frac{(E - \bar{E})^2}{2 \sigma^2 N} \right] \sim
  \frac{1}{\sqrt{N}} \exp \left[ cN - \frac{(E - \bar{E})^2}{2 \sigma^2 N}
  \right],
\end{equation}
where the energy levels are assumed to have Gaussian distribution with mean
$\bar{E}$ and variance $\sigma \sqrt{N}$, and where $\sigma = O (\lambda^4)$.
Then the energy of the ground state $E_0 (\lambda)$ (corresponding to
configuration with the smallest perturbation theory correction) can be
estimated by solving $\rho (E_0) \approx 1$. This implies
\begin{equation}
  E_0 \approx \bar{E} - N \sigma \sqrt{2 c} + O \left( \frac{\log N}{N}
  \right) .
\end{equation}
Notice that this correction is proportional to $N$ rather than $\sqrt{N}$. The
fluctuations of $E_0$ are $O (\sigma)$ and have a Gumbel distribution \cite{textbook}.
This linear scaling is verified numerically in Fig.~\ref{fig:degeneracy} (right) 
where we plot the difference of the largest perturbative correction and the smallest
perturbative correction (so that the common term cancels out) as a function of
$N$.

Next, we show that the level spacing is only $O (\sigma) \equiv O (\lambda^4)$
and does not scale with $N$. Let us compute the probability that the gap to
the first excited state is at least $\Delta$. First, pick a reference energy
$E_{\mathrm{ref}}$ and write down the probability that $\exp (cN) - 1$ levels
have higher energy and 1 level has energy $E_0 = E_{\mathrm{ref}} - \Delta$:
\[ \exp (cN) \left[ 1 - \int_{- \infty}^{E_{\mathrm{ref}}} \ee^{- (E -
   \bar{E})^2 / (2 \sigma^2 N)}  \frac{\dd E}{\sigma \sqrt{2 \pi N}}
   \right]^{\exp (cN) - 1} \frac{1}{\sigma \sqrt{2 \pi N}} \ee^{-
   (E_{\mathrm{ref}} - \bar{E} - \Delta)^2 / (2 \sigma^2 N)} \dd
   E_{\mathrm{ref}} .\]
This expression is non-negligible only if $E_{\mathrm{ref}} - \bar{E} \approx -
N \sigma \sqrt{2 c}$. The desired expression is an integral over $E_{\mathrm{ref}}$,
\begin{equation}
  p (E_1 - E_0 > \Delta) = \int \exp \left( - \frac{\bar{E} -
  E_{\mathrm{ref}}}{\sigma^2 N} \Delta - \frac{\Delta^2}{2 \sigma^2 N} \right) A
  (E_{\mathrm{ref}}) \dd E_{\mathrm{ref}},
\end{equation}
where $A (E_{\mathrm{ref}})$ is a complicated expression independent of $\Delta$.
Replacing $E_{\mathrm{ref}}$ with its approximate value in the exponential and
neglecting the term quadratic in $\Delta$, we obtain
\begin{equation}
  p (E_1 - E_0 \geqslant \Delta) = \exp \left( - \frac{\sqrt{2 c}}{\sigma}
  \Delta \right),
\end{equation}
where we also used the fact that the probability is 1 when $\Delta = 0$. The
same result is obtained for $E_2 - E_1$, $E_3 - E_2$, etc. At the edge of the
spectrum, the spacings are exponentially distributed with mean $\sigma /
\sqrt{2 c}$, i.e. levels have Poisson statistics. These results are not new:
they are well-known in extreme value statistics \cite{textbook} and appear in a solution
of Derrida's random energy model.

\begin{figure}[!ht]
\includegraphics[width=0.6\linewidth]{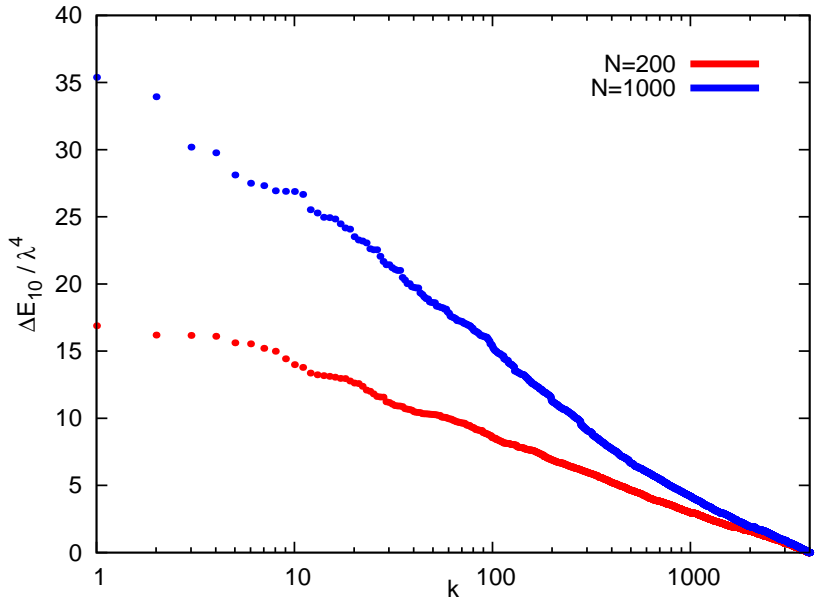}
\caption{The values of normalized level splittings $\Delta E_{10} /  \lambda^4$ for
4000 random instances with $N=200$ and $N = 1000$, sorted in a decreasing order. 
Each dot's $y$-coordinate is the value of the splitting and the $x$-coordinate is its 
index $k$ in the decreasing sequence. A straight line on semilogarithmic plot would
correspond to exponential distribution. Deviation from true exponential is noticeable
for $N=1000$. A clause-to-variable ratio is fixed to $M/N=0.62$.
\label{fig:exp}}
\end{figure}

When a new random clause is added, a fraction of the solutions will
disappear. If we neglect any correlations as before, we can assume that each
solution will satisfy the new clause with finite probability $p < 1$.
Conditioned on the fact that the ``old'' ground state contradicts the new
clause, the probability that old $k$-th excited state satisfies it, but 1st,
2nd, $(k - 1)$-st excited states contradict it,
\begin{equation}
  p_k = p (1 - p)^{k - 1} .
\end{equation}
The gap between old ground and $k$-th excited state $E_k - E_0$ is
distributed with probability density
\begin{equation}
  \rho_k (x) = \left( \frac{\sqrt{2 c}}{\sigma} \right)^k  \frac{x^{k - 1}}{(k
  - 1) !} \ee^{- \frac{\sqrt{2 c}}{\sigma} x} .
\end{equation}
Therefore, the distribution of spacing between the old ground and lowest-lying 
excited state satisfying the new clause is
\begin{equation}
  \rho (x) = \sum_{k = 1}^{\infty} p_k \rho_k (x) = \frac{p \sqrt{2
  c}}{\sigma} \ee^{- \frac{p \sqrt{2 c}}{\sigma} x},
\end{equation}
an exponential distribution with mean $\frac{\sigma}{p \sqrt{2 c}}$. Strictly
speaking, this is not the same as the distribution of the correct quantity
$\Delta E_{10} (\lambda) = E_{\boldsymbol{x}_0}^{(M - 1)} (\lambda) -
E_{\boldsymbol{x}_1}^{(M - 1)} (\lambda)$, where $\boldsymbol{x}_1$ and
$\boldsymbol{x}_0$ correspond to the ground state of instance with $M - 1$ and
$M$ clauses respectively. The addition of new clause introduces a
configuration-dependent correction $O (\lambda^4)$ which is comparable to $O
(\lambda^4)$ level spacing. This means that the levels are somewhat
``reshuffled'', i.e. old ($k + 1$)-st excited state may become smaller in
energy than old $k$-th excited state. We therefore expect that the
distribution of energy differences will deviate from true exponential, but the
characteristic scale should still be $O (\lambda^4)$ with no $N$-dependence.
Fig.~\ref{fig:exp} illustrates the distribution of $\Delta E_{10}$ for a particular
random instance; for large $N$ the distribution still has an exponential tail, but the 
middle of the distribution slightly deviates from true exponential.

We should mention that approximating the distribution of configuration-dependent
4th order corrections can be approximated by a Gaussian only for $E - \bar{E}
\sim \sqrt{N} \lambda^4$, but $E_0$ corresponds to the tail of the
distribution where this approximation is not valid. The probability density of
the sum of $O (N)$ random variables, each having variance $O (\lambda^4)$ is
expected to be exponentially small when we are $O (N \lambda^4)$ away from the
mean; the exact dependence can be computed by considering optimal
fluctuations. Hence, we still expect that $\bar{E} - E_0 \sim N \lambda^4$.
Similarly, level spacing $E_1 - E_0 \sim \lambda^4$, although there is no
guarantee that it is exponential-distributed. Therefore, our conclusions are
independent of this approximation.

Since $\Delta E_{10} \sim \lambda^4$, avoided crossings should not take place
until $\lambda \sim 1$. But for these values of $\lambda$, higher orders of
perturbation theory may not be discarded and the perturbation theory itself
may become divergent, as it should near the quantum phase transition.

Our prediction is in apparent disagreement with the results of numerical
simulations of Ref.~\cite{Alt09} that seem to support the claim that $\Delta E_{10}
\sim \sqrt{N} \lambda^4$. Ref.~\cite{Alt09} correctly examined the edge of the
spectrum: all solutions were enumerated and the 4th order perturbation theory
corrections both before and after adding the new clause were computed for
$\boldsymbol{x}_1$ and $\boldsymbol{x}_0$ that would correspond to the local and
global minima, i.e. having the smallest perturbation theory correction. The average,
median and percentiles of $p \left[ (\Delta E_{10} / \lambda^4)^2 \right]$ as a function
of $N$ for up to $N = 200$ were plotted and a linear fit was found. However,
as we mentioned earlier, for $N \sim 100$ the effects of exponential
degeneracy are not yet prominent. Had the simulation been extended to larger
values of $N$, the flattening of the curves would have been observed suggesting a
finite limit as $N \rightarrow \infty$.

\textbf{Numerical results.} We have extended the numerical study of 
Ref.~\cite{Alt09} to much larger values of $N$. A complete enumeration of all
solutions becomes prohibitively time-consuming as the number of solutions explodes.
However, we are really interested in a solution with the smallest 4th order
perturbation theory correction. From Eq. (\ref{de4}) it is seen that this
correction is linear in binary variables. Finding a solution corresponding to
the ground state is equivalent to solving integer linear programming (ILP)
problem, for which we utilize standard software packages \cite{soft}. ILP algorithms
are more efficient than approaches based on a complete enumeration as entire
branches corresponding to suboptimal solutions are pruned using e.g. LP
relaxations as a lower bound.

\begin{figure}[!ht]
\begin{minipage}{0.6\linewidth}
\includegraphics[width=\linewidth]{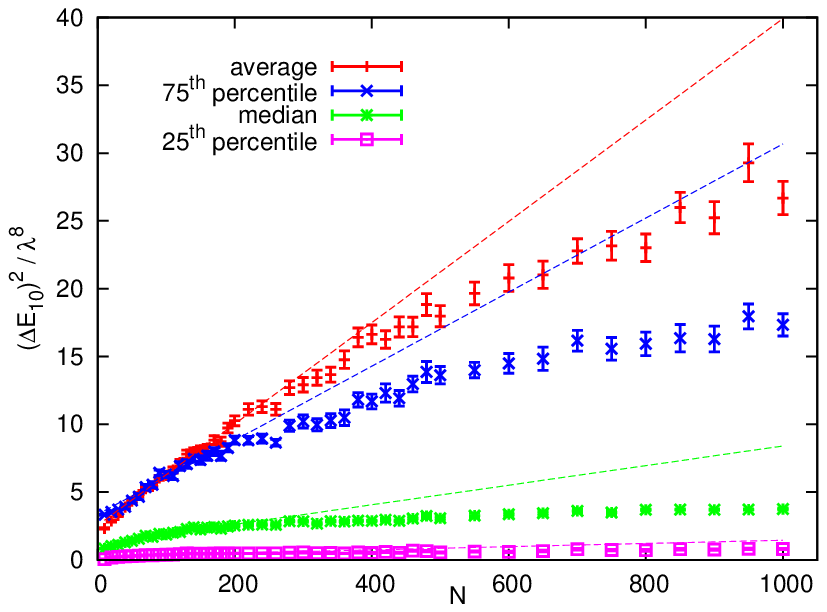}
\end{minipage}
\begin{minipage}{0.35\linewidth}
\includegraphics[width=\linewidth]{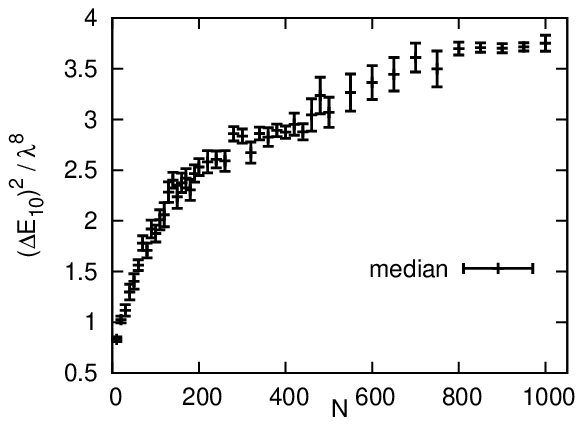}
\includegraphics[width=\linewidth]{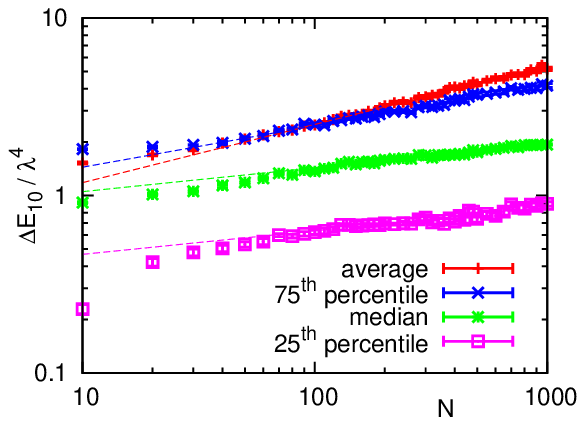}
\end{minipage}
\caption{\emph{Left:} Average as well as 75$^{\mathrm{th}}$, 50$^{\mathrm{th}}$ (median), 
and 25$^{\mathrm{th}}$ percentiles of distribution of $(\Delta E_{10} / \lambda^4)^2$ for different values of $N$
(cf. Fig.~2 from Ref.~\cite{Alt09}). Dashed lines are linear fits for $50 \leqslant N \leqslant 200$.
\emph{Right top:} Just the median of the distribution as a function of $N$.
\emph{Right bottom:} Average and percentiles of $p(\Delta E_{10} / \lambda^4)$ 
(not squared!) on a log-log plot. Dashed lines correspond to power-law fits 
in the interval $100 \leqslant N \leqslant 1000$. The exponents
obtained were: $0.33 \pm 0.01$ (average), $0.23 \pm 0.01$ (75\%), and
$0.13 \pm 0.01$ (median and 25\%); here error estimates refer only to
the goodness of fit. The estimates of the exponents are unreliable since
only one decade of value of $N$ was included in the fit. 
Errorbars correspond to one standard deviation (68\%).
A clause-to variable ratio is $M/N=0.62$.
The results are about 20\% larger than in Ref.~\cite{Alt09}. The discrepancy
might be due to minor difference in the numerical procedure: we chose the added 
clause at random among those that contradict $\boldsymbol{x}_1$, while 
Ref.~\cite{Alt09} restarted from scratch if random clause did not contradict 
$\boldsymbol{x}_1$. The difference is not essential since the the probability 
that a random clause contradicts a particular assignment is finite, but since
this probability depends on the number of ones and zeros in $\boldsymbol{x}_1$,
the distributions are not identical.
\label{fig:num}}
\end{figure}

In Fig.~\ref{fig:num} one can see that the curves are leveling off for larger values of
$N$, in agreement with our argument that $\Delta E_{10}$ should not scale with
$N$. The fact that the average square of the gap and the 75th percentile are
so close to each other for $N \leqslant 200$ (also seen in Ref.~\cite{Alt09}) is not
coincidental. It is an indirect evidence that the distribution of $\Delta E$
is close to exponential since $1 - \ee^{- \sqrt{2}} \approx 0.757$. For larger
$N$, the distribution is not exponential, possibly due to above-mentioned reshuffling 
of energy levels as their density is increased.

The flattening of the curve corresponding to the median is quite pronounced. 
Since average is more sensitive to the tails of the probability distribution, 
even larger values $N$ may be needed to show its approach to the limiting value at 
$N \rightarrow \infty$.

Of course, the present numerical study cannot completely rule out
the possibility that $\Delta E_{10}$ still increases with $N$ with a
power-law exponent smaller than $1/2$. Indeed, the median (which is
more statistically robust measure of scale than the average) seems to
grow as $N^{0.13}$ in the interval $100 \leqslant N \leqslant 1000$.
Tails of the distribution might be responsible for larger exponents
observed for the 75$^{\mathrm{th}}$ percentile and the average.
If the corrections were to grow indefinitely, for sufficiently
large $N$ they would be large enough to cause avoided crossings.
With the assumption that corrections increase as $N^{1/2}$, Ref.~\cite{Alt09}
claims that the mechanism may only set in for very large $N>N_c$,
where the threshold had been estimated as either $N_c \approx 5400$ or
$N_c \approx 86000$ depending on assumptions made. If the corrections
were to rise only as $N^{0.13}$ rather than $N^{0.5}$, the value of $N_c$ would be
pushed even higher. We expect that an observed power-law fit with a finite value of the
exponent is an artifact of using too short an interval (between 100 and 1000). 
An observation that the exponent is close to $1 / \ln 1000$ (corresponding to the
largest size considered) suggests a possibility that corrections increase as a logarithm of $N$.
A logarithmic rise would violate the condition $\lambda_\ast \lesssim 1/\log N$ given 
in Ref.~\cite{Alt09}: indeed, a central point of its argument is the claim
that corrections increase as a finite power of $N$, or much faster than a logarithm.
The less stringent condition conjectured there would be satisfied, but the corresponding 
value of $N_c$ might be astronomically large.

Numerical results clearly contradict the square-root-of-$N$ scaling,
but cannot reliably distinguish an approach to a finite limit from an
extremely slow increase with $N$ (e.g. as a logarithm).
Based on numerical study alone, this scenario cannot be ruled out, but
the theoretical analysis of the previous section, although imprecise,
suggests that the corrections approach a finite limit as $N \to \infty$.
But we can think of no reason that might cause a plausible logarithmic rise.

\begin{figure}
\includegraphics{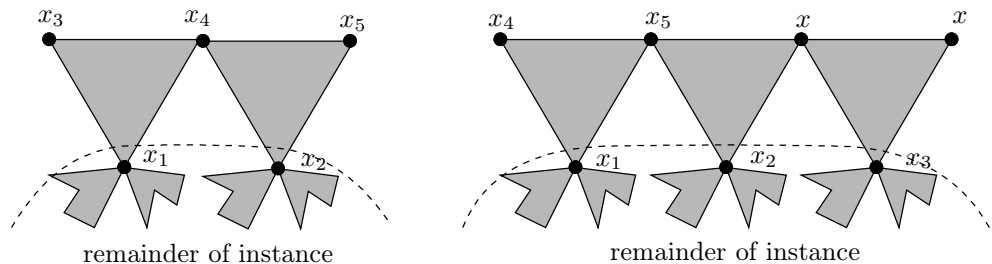}
\caption{\emph{Left:} An example of $O(\lambda^3)$ contribution from the 
degenerate perturbation theory. If $x_1=x_2=0$, two allowed assignments of variables 
$(x_3,x_4,x_5)$: $(0,1,0)$ and $(1,0,1)$ differ by three spin flips.
\emph{Right:} An example of $O(\lambda^4)$ contribution from the degenerate
perturbation theory. $(x_4,x_5,x_6,x_7)$ can be either $(0,1,0,1)$ or
$(1,0,1,0)$ if $x_1=x_2=x_3=0$. The clauses cannot be removed without affecting the satisfiability
of the instance: they prohibit an assignment $x_1=x_3=1$, $x_2=0$.
In each figure solid dots represent binary variables and triangles represent clauses 
in an instance of exact cover problem. Binary variables below the dashed lines 
are involved in other clauses as indicated by zigzag lines.
\label{fig:dPT}}
\end{figure}

\textbf{Concluding remarks.} We want to highlight one important
limitation of the perturbation theory approach. Even for the ``trimmed''
ensemble considered in Ref.~\cite{Alt09}, strictly speaking the largest
configuration-dependent correction is not $O (\lambda^4)$ but rather $O
(\lambda^3)$, the latter coming from degenerate perturbation theory. Indeed,
consider two clauses connected to the remainder of the graph as depicted in
Fig.~\ref{fig:dPT} (left). If both $x_1 = x_2 = 0$ then $(x_3, x_4, x_5)$ can be
assigned either $(0, 1, 0)$ or $(1, 0, 1)$. Since the two configurations with
the same energy differ by 3 bit flips, the splitting caused by the degenerate
perturbation theory causes $O (\lambda^3)$ correction to the energy. It can be
argued that such clauses can be removed: since they can be satisfied for any
value of $x_1$ and $x_2$ they only contribute to trivial degeneracies.
However, in a similar example involving three clauses [see Fig.~\ref{fig:dPT} (right)],
they cannot be removed and yet they contribute $O (\lambda^4)$ due to the
degenerate perturbation theory correction --- the same order as the
correction due to ordinary perturbation theory. In other problems the effect
of degenerate perturbation theory can be stronger: for $K$-SAT it enters as $O
(\lambda)$ correction. The difficulty of dealing with contributions from the
degenerate perturbation theory is a need to diagonalize matrix involving many
solutions. Although ordinary perturbation theory is inadequate, we believe
that our main contention, that $\Delta E$ does not scale with $N$, is still
correct.

The crucial factor in our analysis is the existence of exponentially many
solutions. This phenomenon is common to all combinatorial optimization
problems defined on random hypergraphs. One might ask if in some models
hypergraph ``trimming'' may lift this degeneracy. One such example is
$K$-XOR-SAT problem, where exponential degeneracy can be removed right at the
satisfiability threshold by such trimming. However, perturbative corrections
are independent of bit assignments to all orders of perturbation theory, and
the mechanism described in Ref.~\cite{Alt09} is not applicable there. This is probably
not coincidental: unless local energy landscapes are identical in the vicinity
of all solutions, the exponential degeneracy may not be removed by only
geometric transformations of the random hypergraph.

While we refute the claim that exponentially small gaps appear with high
probability for $\lambda \rightarrow 0$, the general possibility of
exponentially small gaps for finite $\lambda < \lambda_c$ cannot be ruled out.
But estimating the probability of their occurrence might require using
non-perturbative approaches.

We acknowledge the financial support of the United States National Security Agency's 
Laboratory for Physical Sciences. We also acknowledge the support with computational 
resources (32-node Linux cluster) from the  United States Office of Naval Research via 
grant N00014-06-1-0616.

\end{document}